AUTHOR ACCEPTED MANUSCRIPT

# Beyond Curiosity: Habit and Organizational Learning Capability as Predictors of Shadow AI

Nolan Lovett, Educational Leadership & Workforce Development, Old Dominion University (https://orcid.org/0009-0003-1982-830X)

Mohan Yang, Educational Administration and Human Resource Development, Texas A&M University (https://orcid.org/0000-0003-0856-0814)

Correspondence concerning this article should be addressed to Mohan Yang (mohanyang@tamu.edu).

This is the authors' accepted manuscript, the version accepted for publication after peer review. It has not been copyedited, typeset, or proofread by the publisher, and it may differ from the published article. Please cite the published article.

Ethics statement: The study was approved by the Institutional Review Board of Texas A&M University. All participants provided informed consent.

The supplementary materials accepted with the article follow the manuscript.

**Beyond Curiosity: Habit and Organizational Learning Capability**

**as Predictors of Shadow AI**

## Abstract

Generative AI has entered the workplace faster than organizations can govern it. Employees may use AI tools their employer has not sanctioned and conceal that use, but the psychology of this behavior remains poorly understood. We compare an exploratory-disposition account with one centered on reported use and habit. Using survey data from 510 employed adults reporting workplace generative-AI use during the preceding 12 months, we estimated the unique and relative contributions of workplace curiosity, habit, technology-acceptance beliefs, and perceived organizational learning capability through hierarchical regression, bootstrap relative weights, and full-subset dominance analysis. Habit had the largest relative weight, accounting for 28.3% of explained variance, and completely dominated joyous exploration across all 512 possible predictor subsets. Joyous exploration contributed 3.4% and became nonsignificant when use behavior, performance expectancy, and social influence entered the model; habit subsequently explained additional variance. Organizational learning subscales accounted for 24.1% of explained variance; experimentation and knowledge transfer were associated with lower shadow AI, while curiosity-by-learning interactions were nonsignificant. Separate models for unauthorized use and concealment reproduced the habit-dominance pattern. Perceived professional support was associated with lower concealment, although its coefficient did not significantly differ between the two behavioral outcomes. Shadow AI in this sample was more strongly associated with self-reported habit strength and perceived organizational conditions than with exploratory disposition. The cross-sectional design does not establish causal or temporal order.

## 1. Introduction

In April 2023, Samsung discovered that engineers had pasted source code, meeting notes, and chip specifications into ChatGPT (Gurman, 2023). The employees were not acting maliciously but were using a tool to work more efficiently. Within weeks, Samsung banned generative AI, joining JPMorgan Chase, Goldman Sachs, Apple, and other organizations that restricted access after discovering unauthorized use (McGlauflin, 2023). This pattern has since become routine: individuals adopt powerful AI tools on their own initiative, outside sanctioned channels, and often conceal that use. We use shadow AI to encompass unsanctioned AI use for work and concealment of that use (Haag & Eckhardt, 2017; Silic et al., 2025).

Shadow AI has spread across industries. Nationally representative U.S. surveys found that 45% of adults aged 18 to 64 used generative AI as of late 2024. Among employed respondents, 27% had used it for work during the preceding week (Bick et al., 2026). Eloundou et al. (2024) estimated that about 80% of the U.S. workforce could have at least 10% of its tasks affected by large language models. Field experiments help explain why adoption can outpace governance. Generative AI reduced writing-task completion time by 40% and improved output quality (Noy & Zhang, 2023), while Brynjolfsson et al. (2025) documented a 15% average productivity gain among customer-service agents. When a technology delivers immediate performance benefits, employees may adopt it faster than organizations can evaluate, approve, or govern it (Gartner, 2025; UpGuard, 2025).

Shadow AI is, at its core, a human-technology behavior. It concerns how individuals form, sustain, and conceal a pattern of tool use, not the tools alone. Two features make it distinctive within the psychology of technology use. First, generative AI may leave little observable trace. An employee can draft a report with ChatGPT, refine a presentation with Claude, or debug code with Copilot through a personal device and account, producing work whose AI-assisted origins may be difficult to detect. There may be no application on the organization's network and no license in procurement records, leaving governance heavily dependent on users' dispositions, habits, and interpretations of the social environment (Silic et al., 2025). Second, shadow AI can involve concealment as well as unsanctioned use. Many employees manage the visibility of the tool and the AI-assisted nature of their outputs, although unsanctioned use and concealment need not co-occur.

Recent experimental evidence helps explain why concealment is psychologically consequential rather than a peripheral compliance issue. Across four preregistered experiments, Reif et al. (2025) found that AI users anticipated and received negative evaluations of their competence and diligence relative to people using other forms of assistance or no assistance. This evidence positions concealment as a plausible impression-management response to anticipated devaluation, and as a behavior in its own right that adoption-focused models do not address. Shadow AI therefore sits at the intersection of technology use, self-presentation, and the social psychology of work.

Two accounts may explain differences in shadow AI among employees with recent workplace AI use. The first emphasizes individual disposition: curious, exploratory employees may initiate unsanctioned use while experimenting with new tools. This account is better understood as a common intuition than as an established scholarly position. Shadow IT research emphasizes organizational bottlenecks and employees' pursuit of autonomy rather than trait curiosity (Behrens, 2009; Silic & Back, 2014). A related study found that access to GPTs among private-sector competitors increased street-level bureaucrats' intentions to use GPTs against organizational rules (Frisch-Aviram et al., 2025). The intuition nevertheless has a theoretical basis. If shadow AI can function as constructive deviance, joyous exploration is a plausible disposition for initiating it (Kashdan et al., 2018, 2025). The second account centers on habit. Research distinguishes deliberate, intention-driven technology use from habitual use that becomes automatic, cue-triggered, and increasingly decoupled from conscious intention (Ortiz de Guinea & Markus, 2009; Limayem et al., 2007; Venkatesh et al., 2012; Wood & Rünger, 2016). Once use becomes routinized, the factors that initiated experimentation may no longer explain its continuation. If so, the association between curiosity and shadow AI should weaken after established use and habit are modeled.

Exploratory dispositions and habitual AI use do not operate in isolation. Dispositions are expressed within contexts (Johns, 2006), and a learning-supportive work environment may provide sanctioned opportunities to try and discuss new AI tools. We therefore model organizational learning capability (OLC) alongside individual predictors. We test both an additive model, in which OLC is independently associated with shadow AI, and a moderation model, in which the association between curiosity and shadow AI varies with OLC, as situational-strength accounts imply (Meyer et al., 2010).

**The present study.** This study compares two explanations for shadow AI among employees reporting generative-AI use for work during the preceding 12 months. The exploration account locates the behavior in workplace curiosity, especially joyous exploration. The entrenchment account locates it in established use and self-reported habit strength, while allowing organizational learning conditions to explain whether use occurs through legitimate or covert channels. The distinction matters because the accounts imply different governance responses. We therefore compare the unique and relative contributions of curiosity, use behavior, habit, selected UTAUT2 beliefs, OLC, and governance conditions, and examine unauthorized use and concealment both jointly and separately.

1. RQ1. How frequently do employed AI users endorse unauthorized AI use and concealment, and how do they co-occur?
2. RQ2. What is the relative importance of workplace curiosity, self-reported habit strength, and perceived organizational learning capability in explaining shadow AI, and does curiosity contribute unique variance beyond these predictors?
3. RQ3. How do governance configurations (policy and training) relate descriptively to shadow AI?

This study makes three contributions to the psychology of technology use. First, it directly compares an exploratory-disposition account with a habit-based account using bootstrap relative weights and full-subset dominance analysis. Habit accounted for 28.3% of explained variance and completely dominated joyous exploration across all 512 subsets; all four curiosity dimensions fell in the bottom tier. Second, the study treats concealment as a distinct form of visibility management rather than a simple byproduct of unauthorized use. Separate models show the habit-dominance pattern for both unauthorized use and concealment, and perceived professional support is associated with lower concealment, although a direct comparison does not establish that its coefficient differs from that for unauthorized use. Third, the study places organizational learning conditions alongside individual dispositions and established use patterns. Experimentation and knowledge-transfer conditions are associated with lower shadow AI, while curiosity's association attenuates when use behavior, performance expectancy, and social influence enter the model, before habit is added. Together, these findings distinguish correlates of covert use among recent workplace users and motivate longitudinal tests of how initial exploration develops into sustained use.

## 2. Theoretical Background and Hypotheses

**Integrative framework.** The framework assigns a distinct explanatory role to each construct family. Workplace curiosity represents exploratory disposition; use behavior captures the frequency of generative-AI engagement across work tasks; habit captures reported routinization and felt dependence; OLC represents perceived organizational conditions for experimentation and knowledge exchange; social influence represents perceived professional support for AI use; and policy, training, and personal payment describe governance and provision. Unauthorized use and concealment are treated as related but distinct facets, with risk comfort included as an attitudinal indicator and tested separately. The study draws on selected UTAUT2 constructs as competing predictors rather than testing the complete UTAUT2 model.

## 2.1. From shadow IT to shadow AI

Shadow IT refers to the use of technology solutions without formal approval and has long been recognized as a source of cybersecurity and compliance challenges (Haag & Eckhardt, 2017; Silic et al., 2025). It is typically driven by perceived bottlenecks from slow or restrictive IT processes, which prompt employees to adopt alternative tools for higher efficiency (Klotz et al., 2019). Shadow IT research has established that unauthorized technology use is usually productivity-motivated rather than malicious (Silic & Back, 2014), placing it within a broader class of behaviors that management scholars term constructive deviance, namely voluntary rule violations intended to benefit the organization or its stakeholders (Morrison, 2006; Vadera et al., 2013). Case research shows that shadow systems often constitute user-driven innovation that organizations struggle to govern (Györy et al., 2012; Zimmermann et al., 2017), that acceptance of justifications for unsanctioned tools differs between users and nonusers (Haag et al., 2019), and that deviant tools can even generate performance under the right tensions (Haag et al., 2022).

Generative AI introduces features that distinguish shadow AI from traditional shadow IT (Silic et al., 2025). Its outputs are often difficult to trace to a particular tool, its responses may be unpredictable, and it produces new content rather than simply retrieving existing information. These features create what Silic et al. (2025) call a governance drift zone, in which formal policies exist but have limited operational traction. Unauthorized AI tools can also expand organizational threat surfaces through disclosure of sensitive data to third-party models and through compliance violations in regulated industries (Puthal et al., 2025). Practitioner attention has outpaced academic theorizing (Donegan, 2025; Zborowska, 2025), while the peer-reviewed literature remains largely descriptive or governance-focused (Puthal et al., 2025; Silic et al., 2025). Reviews of AI in organizations therefore call for research on how employees actually engage with these tools (Bankins et al., 2024).

### 2.2. Habit and the psychology of sustained technology use

The dominant lens for workplace technology behavior is the acceptance and adoption tradition anchored in the Unified Theory of Acceptance and Use of Technology (UTAUT; Venkatesh et al., 2003) and its consumer extension, UTAUT2 (Venkatesh et al., 2012), which added hedonic motivation, price value, and habit. Shadow AI involves voluntary, consumer-style technology use within organizations. Employees can adopt tools such as ChatGPT, Claude, Copilot, or Gemini through mechanisms similar to those governing consumer applications, and they may pay for access themselves. UTAUT2 therefore provides a useful base. Our theoretical emphasis, however, is on habit, which repeatedly emerges as a strong determinant of continued technology use and often surpasses performance expectancy in UTAUT2 applications (Tamilmani et al., 2021).

Habit is the automaticity that arises from learning and repetition (Triandis, 1980; Venkatesh et al., 2012). In information systems research it has been formalized as a mechanism that limits the predictive power of intention (Limayem et al., 2007), operating through an instant-activation process in which repeated performance lets established attitudes guide behavior automatically. Polites and Karahanna (2012) showed that incumbent-system habit and inertia inhibit new-system acceptance, and Ortiz de Guinea and Markus (2009) argued that habit and emotion together can override the deliberate intention pathway that classical adoption models assume. The broader psychology of habit makes the same point. Habits are activated by recurring context cues and are insensitive to short-term changes in goals, so people often act out of habit even when it conflicts with their intentions (Wood & Rünger, 2016).

For an emerging behavior such as personal generative-AI use, this literature suggests a possible developmental sequence. Whatever first prompts an employee to try a tool, whether curiosity, productivity pressure, or hedonic motivation, may become less relevant as repeated use becomes routinized. Governance strategies directed only at the originating motivation could therefore miss the established behavior. This yields our first hypothesis.

**H1:** Habit is positively associated with shadow AI and contributes more incremental variance than joyous exploration.

Two further UTAUT2 constructs are relevant to concealment and provision. Social influence is theoretically ambiguous in unsanctioned-use contexts. In compliant settings, it predicts adoption (Venkatesh et al., 2003, 2012), but Eckhardt et al. (2009) showed that referent groups can affect adopters and non-adopters differently. When professional referents endorse AI use, the behavior is more socially legitimate, and employees may have less reason to conceal it. We therefore predict a negative association between perceived professional support and shadow AI, particularly concealment. This logic connects to the social-evaluation penalty documented by Reif et al. (2025).

**H2:** Social influence (perceived professional support for AI use) is negatively associated with shadow AI, controlling for use behavior and habit.

Price value extends naturally to a phenomenon distinctive to shadow AI: the employee who pays out of pocket. Personal payment represents an individual investment in access to AI tools. Such spending may signal unmet organizational provision (Klotz et al., 2019), although it may also reflect product preference or unusually strong adoption commitment. We therefore expect personal payment to predict elevated shadow AI after adjustment for use frequency.

**H3:** Employees who personally pay for AI subscriptions report elevated shadow AI, controlling for use behavior and habit.

### 2.3. Workplace curiosity as an exploration disposition

While UTAUT2 explains how performance expectations and habit sustain use, it largely overlooks dispositions that may prompt initial, unauthorized experimentation with AI. Hedonic motivation captures situational pleasure in using a technology (Venkatesh et al., 2012), whereas curiosity captures a trait-level disposition to seek novelty, ideas, and stimulation (Kashdan et al., 2018). Curiosity motivates exploratory behavior, information seeking, and learning (Lievens et al., 2022); it also supports creative problem solving and is actively cultivated within organizations (Hardy et al., 2017; Kashdan et al., 2023; Slavich et al., 2026).

Curiosity is not value-neutral with respect to organizational rules. Constructive deviance has three defining features: it is voluntary, violates a significant organizational norm, and is enacted with pro-organizational rather than malicious intent (Morrison, 2006; Vadera et al., 2013). Shadow AI can satisfy these conditions when employees use unapproved tools to improve their work (Silic & Back, 2014), placing it alongside bootlegging and other innovation-motivated rule breaking (Dahling & Gutworth, 2017; Globocnik et al., 2022; Gill, 2026). Curiosity therefore supplies a plausible dispositional account of initial experimentation.

Curiosity has distinct dimensions in general and workplace settings (Kashdan, Disabato, et al., 2020; Kashdan, Goodman, et al., 2020). The Multidimensional Workplace Curiosity Scale (M-WCS; Kashdan, Goodman, et al., 2020), whose brief form has shown generalizability across nine countries (Kashdan et al., 2025), captures four work-relevant dimensions: Joyous Exploration, Deprivation Sensitivity, Stress Tolerance, and Openness to People's Ideas. The exploration account implies that joyous exploration should be positively associated with shadow AI. We also expect openness to people's ideas to be negatively associated with shadow AI because employees who actively value others' perspectives may be more likely to surface AI use and absorb signals about appropriate use.

**H4a**: Joyous exploration is positively associated with shadow AI at the bivariate level and in models that exclude habit and use behavior.

**H4b:** Openness to People's Ideas is negatively associated with shadow AI.

### 2.4. The organizational learning environment as context

Individual curiosity and habitual use are shaped by the learning environment of the enterprise. Organizational learning capability (OLC) refers to the managerial and organizational characteristics that enable and promote learning, conceptualized as antecedent conditions rather than learning outcomes (Chiva et al., 2007; Goh & Richards, 1997; Tohidi et al., 2012). The construct draws on the distinction between exploration and exploitation (March, 1991) and on instrument-development work concerning learning capability (Goh & Richards, 1997; Jerez-Gómez et al., 2005). Tohidi et al. (2012) synthesized earlier instruments and validated dimensions including experimentation, risk taking, and knowledge transfer, each associated with innovation across sectors (Alegre & Chiva, 2008; Chiva & Alegre, 2009).

OLC is the organizational-design counterpart to governance arrangements such as policies, training, and approval workflows. Silic et al. (2025) document that employees can access generative AI through personal devices outside organizational visibility even where restrictive policies exist, while Limayem et al. (2007) show that habit can weaken the role of deliberate intention after behavior becomes established. Learning-supportive environments may offer legitimate channels for experimentation and knowledge exchange. Less supportive environments may leave employees with fewer sanctioned ways to meet emerging task needs. We therefore model OLC as a set of direct, additive correlates of shadow AI. Situational-strength theory (Meyer et al., 2010) raises a separate possibility, that strong learning environments weaken the association between disposition and behavior; we probe that multiplicative form in a supplementary analysis.

**H5:** OLC dimensions are negatively associated with shadow AI.

**H6:** OLC dimensions collectively account for more variance in shadow AI than workplace curiosity dimensions in relative weights analysis.

If established use and habit are more consequential than exploratory disposition, the positive association between joyous exploration and shadow AI should diminish once use behavior and habit are entered. This is a statistical prediction about attenuation; the cross-sectional design cannot establish a developmental sequence.

**H7:** The positive association between joyous exploration and shadow AI will attenuate substantially after use behavior and habit are included in the model.

These hypotheses compare the unique and relative contributions of curiosity, established use, habit, and learning conditions among employees with recent workplace AI use. A substantial curiosity coefficient after adjustment would favor an exploration account of differences among these users; larger contributions from use behavior and habit would favor the entrenchment account. This comparison does not distinguish initial adoption from subsequent habit formation, and indirect-association models remain exploratory. Figure 1 summarizes the framework.

Figure 1. Integrated conceptual model and analytic sequence.

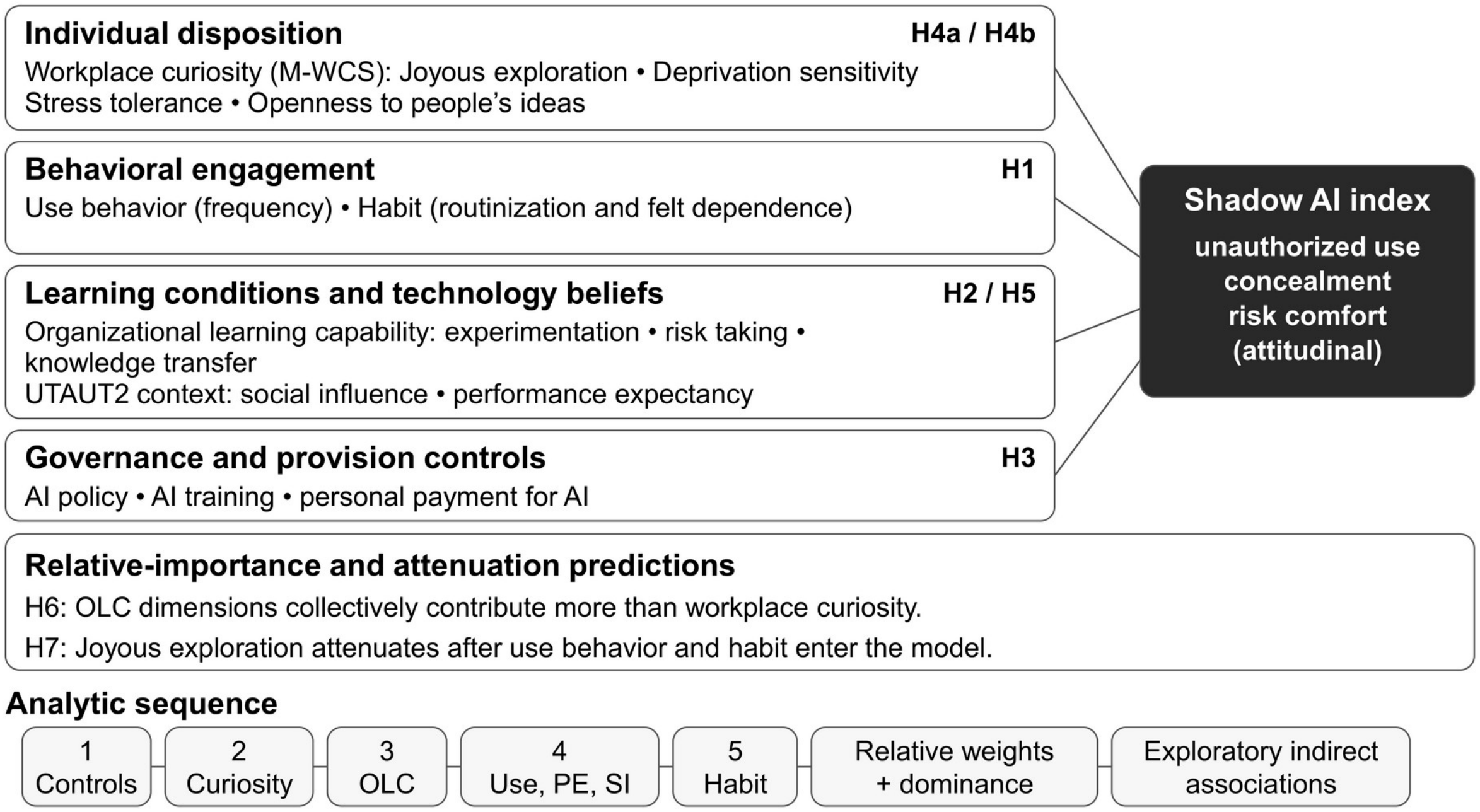

## 3. Method

**Research design.** This theory-driven study used a cross-sectional, quantitative survey and treated the individual employee as the unit of analysis. Directional hypotheses were evaluated deductively; the study and analysis plan were not preregistered. Prevalence estimates and policy-training configurations were descriptive, while robustness checks and indirect-association models were treated as sensitivity or exploratory analyses. Recruitment used a nonprobability online panel, and the focal analysis was restricted to employed adults who reported workplace generative-AI use during the preceding 12 months. The design permits comparison of individual, behavioral, organizational, and governance predictors, but it does not establish causal or temporal order.

### 3.1. Data and sample

Data were collected on 26–27 August 2025 through Prolific, which has demonstrated comparatively strong data quality in online human-subjects research (Douglas et al., 2023; Peer et al., 2022). The study page directed participants to a university-managed Qualtrics survey. The survey included two embedded attention checks. Participants were paid a fixed US$4.00 for a survey advertised to take approximately 20 minutes, equivalent to a planned rate of US$12.00 per hour. Prolific displayed an effective hourly reward rate of US$14.55 based on its median completion time of 16 minutes 30 seconds. The raw export contained 570 response records. We excluded one consent decline, 12 records that failed one or both attention checks, and six additional records with completion times below one third of the sample median. This left 551 quality-screened responses, all of which passed both attention checks. The focal analyses were restricted to respondents who reported using generative AI tools for work during the preceding 12 months. Of the screened respondents, 510 met this criterion and had complete data on the focal regression variables; the remaining 41 did not meet the focal workplace-use criterion and were not included in the focal analyses. Among the quality-screened responses, minimum completion time was 6.1 minutes and the median was 16.2 minutes. The instrument covered a broader set of technology-acceptance and workplace constructs than analyzed here; the present study draws on the subset relevant to its research questions. The questionnaire as administered, including item wording and response options, is provided in the separate Supplementary Questionnaire file. Supplementary

Table S2 reports a selection audit comparing focal users with screened respondents who did not meet the workplace-use criterion. Information on nonrespondents and suitable population benchmarks was unavailable, so this comparison evaluates selection into the focal workplace-use group and does not establish the absence of nonresponse bias.

The focal sample averaged 39.2 years old (SD = 10.2, range 20 to 65) and had 11.1 years of work experience on average (SD = 8.5). Gender composition was 60% male, 39% female, and 1% nonbinary. Industries included technology (21%), professional services (13%), healthcare (12%), retail (11%), education (9%), manufacturing (9%), financial services (8%), and other sectors (17%). Among 506 respondents with observed residence, 48.6% lived in the United States, 24.3% in the United Kingdom, 9.5% in Canada, 7.1% in Mexico, and 10.5% in other countries. Four focal respondents did not report residence. This multinational composition broadens geographic coverage relative to a single-country sample, but the opt-in panel does not support population-representative claims. Regarding organizational context, 35% reported a formal AI policy, 38% reported organizational AI training, and 50% reported self-initiated AI training. The most commonly selected tool offerings were ChatGPT and Gemini. The study was approved by the Institutional Review Board of one of the authors' institutions. Participants provided electronic informed consent and were informed that their responses would be kept confidential and that reports and publications would not identify them. Prolific did not disclose participants' identities to the researchers and supplied pseudonymized platform IDs. The external Qualtrics survey also included optional follow-up contact fields for respondents who volunteered for further contact; those fields and technical metadata were excluded from the analytic data and retained only in the restricted raw file. The survey data used in this article were analyzed without names or direct identity information and were reported only in aggregate.

### 3.2. Measures

#### *3.2.1. Shadow AI behaviors*

We identified no validated instrument designed specifically to assess shadow AI. Prior shadow-IT scales primarily capture unauthorized use and do not jointly measure concealment or comfort with unsanctioned risk, so we developed a provisional shadow-AI index for this study. Three items were rated on 7-point scales (1 = strongly disagree, 7 = strongly agree): unauthorized use (I sometimes use AI tools in ways not explicitly approved by my organization), concealment (I have strategies to avoid detection of my AI use), and risk comfort (I am comfortable taking risks by using AI tools that my organization has not sanctioned). The first two items represent distinct behavioral facets consistent with Haag and Eckhardt's (2017) definition. Risk comfort is attitudinal, so the three-item index is treated as provisional and the behavioral facets are also analyzed separately. The index is an additive, unweighted mean: a higher score does not require endorsement of all three items, and use that is not explicitly approved is not necessarily known to be prohibited. It showed acceptable internal consistency ($\alpha$ = .76); inter-item correlations ranged from .48 to .55, and standardized loadings in the full measurement model ranged from .70 to .75. The composite also related to disclosure behavior and personal payment in the expected directions. Because the topic is sensitive, underreporting is possible. The percentages describe item endorsement, not independently verified conduct. We therefore foreground separate models for unauthorized use and concealment and report a two-item behavior-only sensitivity analysis in Section 4.4. A separate single item measured security consciousness and was analyzed independently because it correlated negatively with the shadow-AI items. Respondents also reported to whom and how they disclose AI use and rated six influences on disclosure: organizational policy and guidelines, task type, audience or stakeholder, anticipated reactions from others, professional norms and standards, and personal comfort.

#### *3.2.2. Workplace curiosity*

Twelve items, three per dimension, from the brief form of the M-WCS (Kashdan, Goodman, et al., 2020; nine-country validation in Kashdan et al., 2025) assessed four dimensions on 5-point scales (1 = very slightly or not at all, 5 = extremely): joyous exploration (α = .86, M = 3.80), deprivation sensitivity (α = .87, M = 3.56), stress tolerance (α = .83, M = 3.37), and openness to ideas (α = .89, M = 4.25).

#### *3.2.3. Habit*

Three UTAUT2 items (Venkatesh et al., 2012) assessed habitual AI use on 7-point scales: the use of GenAI has become a habit for me, I feel addicted to using GenAI, and I feel I must use GenAI (α = .85, M = 3.41). These items capture both automaticity and felt compulsion. The automaticity item (M = 4.50) was endorsed substantially more than the compulsion items (M = 2.71 and 3.00), indicating higher average endorsement of routine use than of felt compulsion. In a single-factor measurement model, however, the compulsion items loaded more strongly (λ = .87 and .91) than the automaticity item (λ = .68), so we interpret the latent construct as self-reported habit strength with a dependency component rather than pure context-cued automaticity (Wood & Rünger, 2016). Section 4.5 reports item-level robustness checks, and the Limitations discuss instruments built specifically for automaticity.

#### *3.2.4. Use behavior*

Five items measured task-specific AI use frequency on 7-point scales (1 = never, 7 = multiple times per day) across five task domains: information processing, content creation and refinement, analysis and problem-solving, ideation and brainstorming, and automation and enhancement (α = .84, M = 4.20). Habit and use behavior are distinct in UTAUT2. Use captures frequency, and habit captures automaticity and dependence.

#### *3.2.5. Performance expectancy and social influence*

Perceived usefulness of AI tools (α = .92, M = 5.67) and the degree to which professionally important others support AI use (α = .94, M = 4.60) were each assessed with three UTAUT2 items on 7-point scales.

#### *3.2.6. Organizational learning capability*

Three subscales from validated organizational learning measures (Tohidi et al., 2012) on 5-point scales captured experimentation (α = .85, 4 items, M = 3.45), risk taking (α = .87, 3 items, M = 2.91), and knowledge transfer (α = .85, 5 items, M = 3.94). Two additional subscales were examined and excluded. Managerial commitment contained reverse-worded items that, even after correction, retained a low CFA loading and produced worse fit for a four-factor than a three-factor model. External interaction showed inadequate reliability (α = .55). Both subscales were excluded from the analytic model. The retained scales describe respondents' perceived learning conditions; they are not independently verified organization-level measurements.

#### *3.2.7. Controls*

Analyses controlled for age, gender, years of work experience, organizational AI policy, organization-provided AI training, and personal payment for AI tools. Age and experience were mean-centered. Gender was coded 1 for male and 0 for the other reported categories; policy was coded 1 for present and 0 for no or unsure; training was coded 1 for organization-provided training and 0 otherwise; and personal payment was coded 1 when respondents selected the personally funded option. Personal payers represented 16.9% of the sample and reported a median expenditure of $20 per month.

### 3.3. Analytic strategy

Confirmatory factor analysis (CFA) evaluated the measurement model before substantive analysis. A 10-factor model across the 33 reflective predictor items, estimated with robust maximum likelihood (MLR), fit well, $\chi^2(450) = 792.37$, $p < .001$, robust CFI = .964, TLI = .958, RMSEA = .041 [.037,

.046], SRMR = .041 (Hu & Bentler, 1999). The four-factor M-WCS fit well and outperformed a one-factor alternative. All standardized loadings exceeded .58, with most above .70. An extended 12-factor model adding five use-behavior items and three shadow-AI items, estimated with maximum likelihood (ML), also fit acceptably, $\chi^2(713) = 1400.20$, CFI = .948, TLI = .940, RMSEA = .043, SRMR = .043. Shadow-AI loadings ranged from .70 to .75. Supplementary Table S5a reports the full-sample CFA fit statistics, including the four-factor and one-factor M-WCS models. Table S5d reports standardized loadings for all 41 items in the extended 12-factor ML model. Country comparisons excluded four missing residence responses, leaving United States and United Kingdom (n = 369) versus other countries (n = 137). Separate ML tests supported equal loadings for 11 of 12 scales; stress tolerance showed a significant decrement, $\Delta\chi^2(2) = 6.61$, $p = .037$. The simultaneous 12-factor test also rejected exact loading equality, $\Delta\chi^2(29) = 50.64$, $p = .008$, although the change in CFI was small (−.002). Measurement equivalence is therefore mixed rather than established for every scale; Supplementary Tables S5b and S5c report the two-group and four-country checks. The shadow-AI factor was distinguishable from habit and use behavior: heterotrait-monotrait ratios were .42 and .37, respectively, below .85 (Henseler et al., 2015); average variance extracted (.52) exceeded its squared factor correlations with habit (.16) and use behavior (.13) (Fornell & Larcker, 1981); and merging shadow AI with either construct worsened fit, $\Delta\chi^2(11) = 382.01$ and 411.04, respectively, both $p < .001$.

Because all focal constructs were self-reported, we conducted post hoc diagnostics for common method variance (Podsakoff et al., 2003). For the 33 reflective predictor items, the first unrotated principal component accounted for 31.02% of variance. A single-factor MLR model for these same items fit poorly (CFI = .403, TLI = .363, RMSEA = .161, SRMR = .137). For the extended 41-item model, an unmeasured latent method-factor sensitivity model estimated with ML improved fit (CFI = .963, RMSEA = .037) and attributed an average of 16.5% of indicator variance to the method factor. These diagnostics show that a single factor does not adequately reproduce the item covariance structure; they do not establish that substantive coefficients are free of common method bias.

Pre-regression diagnostics revealed no focal variable with skewness beyond |2| or kurtosis beyond |7| (Curran et al., 1996). All variance inflation factors were below 3.0 (maximum = 2.69). The Breusch–Pagan test indicated heteroscedasticity ($p < .001$), motivating HC3 robust standard errors (Long & Ervin, 2000). The original Ramsey RESET test using squared and cubed fitted values was significant ($p = .023$); an HC3 version also rejected the linear specification ($p = .007$). Individual quadratic checks identified curvature for habit ($B = 0.06$, HC3 $p = .019$, $\Delta R^2 = .008$). We retain the original linear comparison but interpret it as an approximation rather than evidence of strictly linear associations. The Durbin–Watson statistic was 2.03, and the maximum Cook's distance was .024. These diagnostics cannot establish independence or exclude all influential-case concerns in an opt-in survey.

Analysis proceeded in four phases. Phase 1 used a five-step hierarchical ordinary least squares regression with HC3 robust standard errors, using residual-degrees-of-freedom t tests and confidence intervals for coefficients, and conventional nested-model F tests for block increments. Blocks entered controls, curiosity dimensions, OLC subscales, performance expectancy plus use behavior and social influence, and habit. This sequence placed curiosity early as the focal account under test and entered habit last as its strongest theoretical competitor. Because some steps introduced multiple predictors, step-level increments cannot be attributed to individual variables; the relative-importance analyses provide that decomposition. Phase 2 used Johnson relative weights with 5,000-resample bootstrap confidence intervals, general dominance weights, and full-subset dominance analysis (Budescu, 1993; Johnson, 2000; Tonidandel & LeBreton, 2011). The full model including controls explained 29.7% of variance; relative-importance percentages decompose the 24.5% explained by the 11 substantive predictors alone. Phase 3 decomposed the outcome by estimating parallel final-step models and relative weights for unauthorized use and concealment, and probed curiosity-by-OLC moderation. Phase 4 comprised prevalence, personal-payer effect sizes, policy-training configurations, disclosure-factor analyses, and exploratory indirect-association models.

The analyses were executed in R 4.5.1 on Windows x64. The reproduction package preserves the original analysis specifications, applies the corrected Johnson relative-weight function, and adds documented verification scripts using lm, sandwich, lmtest, boot, psych, and lavaan. The package records fixed resampling seeds, package versions, session information, source hashes, and machine-readable outputs. Its run_all.R entry point uses relative paths and deidentified inputs; a separate audit variable preserves missing country responses rather than recoding them as other countries. The deidentified data, codebook, scripts, and verification outputs are available from the corresponding author on reasonable request. Supplementary Table S3 maps package contents and execution and explains privacy protections; Supplementary Table S4 reports final-model confidence intervals.

## 4. Results

### 4.1. Prevalence and co-occurrence (RQ1)

Among the 510 respondents reporting workplace AI use during the preceding 12 months, 22.2% at least somewhat agreed (scores of 5 or higher on the 7-point scale) that they sometimes use AI tools not explicitly approved by their organization, and 21.4% reported having strategies to avoid detection (11.8% and 11.6%, respectively, selected agree or strongly agree). An additional 32.2% endorsed comfort taking risks with unsanctioned AI tools. The correlation between unauthorized use and concealment was moderate ($r = .52$, $p < .001$), suggesting related but distinct behaviors. Fifty-five respondents (10.8%) endorsed both items at a score of 5 or higher.

Disclosure patterns were more nuanced than the prevalence figures suggest. Most respondents disclosed AI use to their team (69.4%) and direct supervisor (65.7%), though 13.7% told no one. Among respondents answering the disclosure-method question ($n = 440$), 62.5% openly acknowledged AI use, while 8.4% reported presenting AI-assisted work as fully their own. Concealment may be selective:

employees could acknowledge AI use generally while concealing particular uses. The aggregate percentages do not establish that pattern within individuals, but it is consistent with concealment as a possible impression-management response (Reif et al., 2025).

**4.2. Hierarchical regression (RQ2)**

Table 1 presents the five-step hierarchical regression. Step 1, containing controls only, explained 9.2% of variance. Training was negatively associated with shadow AI (B = −0.46, p = .012), and personal payment was positively associated (B = 0.92, p < .001). Age, gender, experience, and policy presence were nonsignificant. In Step 2, curiosity dimensions added significant incremental variance ($\Delta R^2$ = .026, p = .006). Joyous exploration was positively associated with shadow AI (B = 0.29, p = .008), while openness to people's ideas was negatively associated (B = −0.25, p = .031), supporting H4a and H4b at this stage. In Step 3, OLC added further variance ($\Delta R^2$ = .039, p < .001). Experimentation (B = −0.31, p = .014) and knowledge transfer (B = −0.24, p = .046) were negatively associated with shadow AI, while risk taking was nonsignificant (B = 0.09, p = .293), providing partial support for H5. Joyous exploration remained significant and increased slightly (B = 0.33, p = .003), indicating that OLC did not account for its association at this step.

The focal change occurred at Step 4. When performance expectancy, use behavior, and social influence entered as a block ($\Delta R^2$ = .103, p < .001), the joyous-exploration coefficient declined from B = 0.33 (p = .003) to B = 0.13 (p = .185), while use behavior (B = 0.36, p < .001) and social influence (B = −0.16, p = .008) were significant. Adding habit at Step 5 explained a further 3.6% of variance (p < .001). Habit was positively associated with shadow AI (B = 0.24, p < .001), and joyous exploration remained nonsignificant (B = 0.09, p = .362). The full model explained 29.7% of variance. Experimentation (B = −0.36, p = .002), knowledge transfer (B = −0.22, p = .047), training (B = −0.46, p = .009), personal payment (B = 0.50, p = .007), performance expectancy (B = 0.18, p = .014), use behavior (B = 0.26, p < .001), and social influence (B = −0.18, p = .002) remained significant. H1, H2, and H3 were supported;

H4a was supported at its specified bivariate and early-model stages; H4b was supported only before the use-pattern block; and H5 received partial support. H7 was supported as a regression prediction: joyous exploration attenuated when performance expectancy, use behavior, and social influence entered, and remained nonsignificant after habit was added. This sequence does not establish temporal mediation.

*[Insert Table 1 about here]*

Table 2 presents descriptive statistics and bivariate correlations. Shadow AI was positively associated with habit ($r = .33$, $p < .001$), use behavior ($r = .30$, $p < .001$), performance expectancy ($r = .21$, $p < .001$), and joyous exploration ($r = .12$, $p = .007$), and negatively associated with OLC experimentation ($r = −.17$, $p < .001$) and knowledge transfer ($r = −.18$, $p < .001$). Social influence showed a near-zero bivariate correlation ($r = −.02$, ns), consistent with the multivariate suppressor effect. All variance inflation factors were below 3.0.

*[Insert Table 2 about here]*

**4.3. Relative importance analysis (RQ2)**

Table 3 presents Johnson relative weights and raw-weight bootstrap 95% confidence intervals for the 11 substantive predictors (5,000 resamples; controls excluded, total $R^2 = .245$). Habit ranked first, accounting for 28.3% of explained variance, followed by use behavior (21.7%), experimentation (12.5%), performance expectancy (11.9%), and knowledge transfer (9.8%). Joyous exploration contributed 3.4%; deprivation sensitivity, stress tolerance, and openness to people’s ideas each contributed less than 2%. OLC collectively accounted for 24.1%, compared with 7.5% for curiosity, supporting H6; the grouped contrast favored OLC in 98.8% of bootstrap resamples. General dominance weights, which average incremental contributions across all subset sizes, also ranked habit first (28.0%), use behavior second (22.7%), and joyous exploration seventh (2.7%). Bootstrap intervals overlap for the smaller weights, so their exact ordering is uncertain. Habit ranked first in 74.1% of resamples, and its raw-weight difference from joyous exploration was .061 (95% CI [.027, .100]). Joyous exploration fell in the bottom five in 80.5% of resamples but ranked last in only 0.4%; the evidence supports a comparatively small contribution rather than a uniquely last-place rank.

Dominance analysis formally tested the joyous-exploration-versus-habit comparison. Habit completely dominated joyous exploration, contributing more incremental variance at every subset size across all 512 possible subsets of the remaining nine predictors (average $\Delta R^2$ = .038 to .112 for habit versus .002 to .017 for joyous exploration). Complete dominance is the strongest form of relative-importance evidence (Budescu, 1993). None of the predictor subsets examined gave joyous exploration a larger incremental contribution than habit, and habit's relative weight was about eight times as large. The association between joyous exploration and shadow AI became nonsignificant when performance expectancy, use behavior, and social influence entered the hierarchical model, before habit was added.

*[Insert Table 3 about here]*

### 4.4. Decomposing the composite: unauthorized use and concealment

Because the composite averages conceptually distinct facets, we re-estimated the final-step model and the relative-importance analyses separately for each item (Table 4). The habit-dominance pattern replicated for both behavioral facets. For unauthorized use (model $R^2$ = .207), habit retained the largest relative weight (B = 0.24, $p < .001$), accounted for 26.1% of substantive explained variance, and completely dominated joyous exploration across all 512 subsets. For concealment ($R^2$ = .236), habit (B = 0.29, $p < .001$; 36.2%) and use behavior (30.8%) together carried two thirds of the substantive explained variance, and complete dominance over joyous exploration again held in all 512 subsets. Joyous exploration uniquely predicted neither outcome ($p$ = .072 and $p$ = .773).

Social influence was negatively associated with concealment (B = −0.20, $p$ = .011), while the unauthorized-use coefficient was nonsignificant (B = −0.12, $p$ = .116). A paired outcome contrast using the same respondents and predictors did not establish a difference between these coefficients: B = −0.08, HC3 SE = 0.09, $t(492) = -0.82$, $p$ = .415, 95% CI [−0.26, 0.11]. The results therefore do not establish a concealment-specific association. Policy presence, nonsignificant for the composite, was associated with higher concealment ($B = 0.42$, $p = .046$); this isolated result does not show that restrictions caused concealment. OLC accounted descriptively for 30.0% of substantive explained variance in unauthorized use and 9.8% in concealment, a comparison of model decompositions rather than a test of coefficient differences. Risk comfort was associated with habit ($B = 0.21$, $p = .002$), personal payment ($B = 0.72$, $p = .003$), and training ($B = −0.61$, $p = .009$). The behavior-only mean retained the core habit–curiosity pattern (habit $B = 0.26$, $p < .001$; joyous exploration $B = 0.14$, $p = .197$). Some secondary results differed: policy was significant for this two-item outcome ($p = .040$), compared with $p = .192$ for the three-item index.

*[Insert Table 4 about here]*

**4.5. Moderation probe and robustness checks**

Situational-strength accounts imply that learning-supportive environments could weaken the association between exploratory disposition and covert use. We tested this multiplicative form by adding three mean-centered joyous-exploration-by-OLC product terms to the final model. The HC3 joint Wald test was nonsignificant, $F(3, 489) = 1.18$, $p = .319$ ($\Delta R^2 = .006$). A sensitivity block crossed all four curiosity dimensions with all three OLC subscales; its HC3 joint test was also nonsignificant, $F(12, 480) = 0.66$, $p = .786$ ($\Delta R^2 = .014$). These tests did not detect moderation. Nonsignificance does not establish equivalence or rule out smaller interactions, and the data support an additive description without demonstrating that the underlying relationships are exclusively additive.

Further checks probed measurement and data quality. Replacing the habit scale with its automaticity item left that item significant ($B = 0.09$, $p = .045$) and joyous exploration nonsignificant ($B = 0.11$, $p = .257$); the compulsion pair had a larger coefficient ($B = 0.24$, $p < .001$). The result therefore concerns self-reported habit strength, including a substantial dependency component. Adding an observed-residence indicator, United States and United Kingdom versus other countries, to the 506 complete-country cases retained the core habit–curiosity pattern; the country coefficient was nonsignificant (B = 0.18, p = .238). Measurement equivalence remained mixed, as detailed in Section 3.3 and Supplementary Tables S5b and S5c. The method-factor diagnostics also cannot exclude common method bias. Finally, no retained respondent completed the survey in less than one third of the quality-screened median; excluding the 12 fastest focal respondents, all under eight minutes, retained the core habit–curiosity conclusion.

**4.6. Governance and personal payment (RQ3)**

Employees who personally paid for AI subscriptions, 16.9% of the sample, reported higher shadow-AI scores across unauthorized use (d = 0.43), concealment (d = 0.52), and risk comfort (d = 0.65). Personal payment remained associated with shadow AI in the full model (B = 0.50, p = .007) after adjustment for curiosity, OLC, use behavior, habit, and social influence. This pattern may signal unmet organizational provision, but it does not establish why employees paid. Alternative explanations include early-adopter commitment and preferences for premium, private, or otherwise differentiated tools.

Table 5 summarizes respondents' reports of policy and training; the groups do not represent independently sampled organizations. Omnibus ANOVAs were nonsignificant for unauthorized use, F(3, 506) = 1.36, p = .253, and concealment, F(3, 506) = 1.34, p = .262. Descriptively, the policy-only group (n = 48) had the highest endorsement rates (33.3% and 31.2%), and the training-only group (n = 65) the lowest (20.0% and 13.8%); the neither (n = 268) and both (n = 129) groups fell between them. Configuration differences were significant for risk comfort, F(3, 506) = 6.26, p < .001, and security consciousness, F(3, 506) = 4.03, p = .007. The omnibus tests do not identify every pairwise difference. Security-consciousness means also do not support a general claim that training recipients scored higher regardless of policy: the policy-only mean was 6.02, compared with 5.97 for respondents reporting both policy and training.

*[Insert Table 5 about here]*

When asked how influential each of six factors was in their decision to disclose AI use, respondents rated personal comfort (M = 5.37) and task type (M = 5.16) highest, and organizational policy and guidelines (M = 4.07) and anticipated reactions from others (M = 4.00) lowest. Entered as a block over the full final-step model, the six ratings improved prediction of concealment ($\Delta R^2 = .027$, $p = .007$) but not unauthorized use ($p = .212$). Anticipated reactions from others was the only unique predictor of concealment ($B = 0.19$, $p < .001$). Thus, a factor respondents rated as relatively less influential still carried unique statistical information about concealment.

**4.7. Exploratory attenuation and indirect associations**

Supplementary bootstrap models examined whether attenuation was consistent with indirect associations (5,000 resamples). In the simple habit model, the fitted indirect association was 0.20 (percentile 95% CI [0.13, 0.27]), the total association was $c = 0.22$ ($p = .007$), and the direct association adjusted for habit was $c' = 0.02$ ($p = .82$). In the serial model, the indirect associations through use behavior alone, habit alone, and use behavior followed by habit had nonzero bootstrap intervals; the fitted direct association was −0.09 (95% CI [−0.24, 0.07]). These supplementary specifications differ from the fully adjusted hierarchical model and are reported separately in the reproduction outputs. They do not establish mediation or temporal order (Maxwell & Cole, 2007). Figure 2 displays all predictor coefficients and HC3 confidence intervals from the full Step 5 model.

Figure 2. Full Step 5 model predicting the shadow-AI composite. Points are unstandardized coefficients; bars are 95% HC3 t confidence intervals with 492 residual degrees of freedom. N = 510. Every coefficient comes from the same model, adjusted for all other displayed predictors. Coefficient magnitudes across differently scaled predictors are not measures of relative importance.

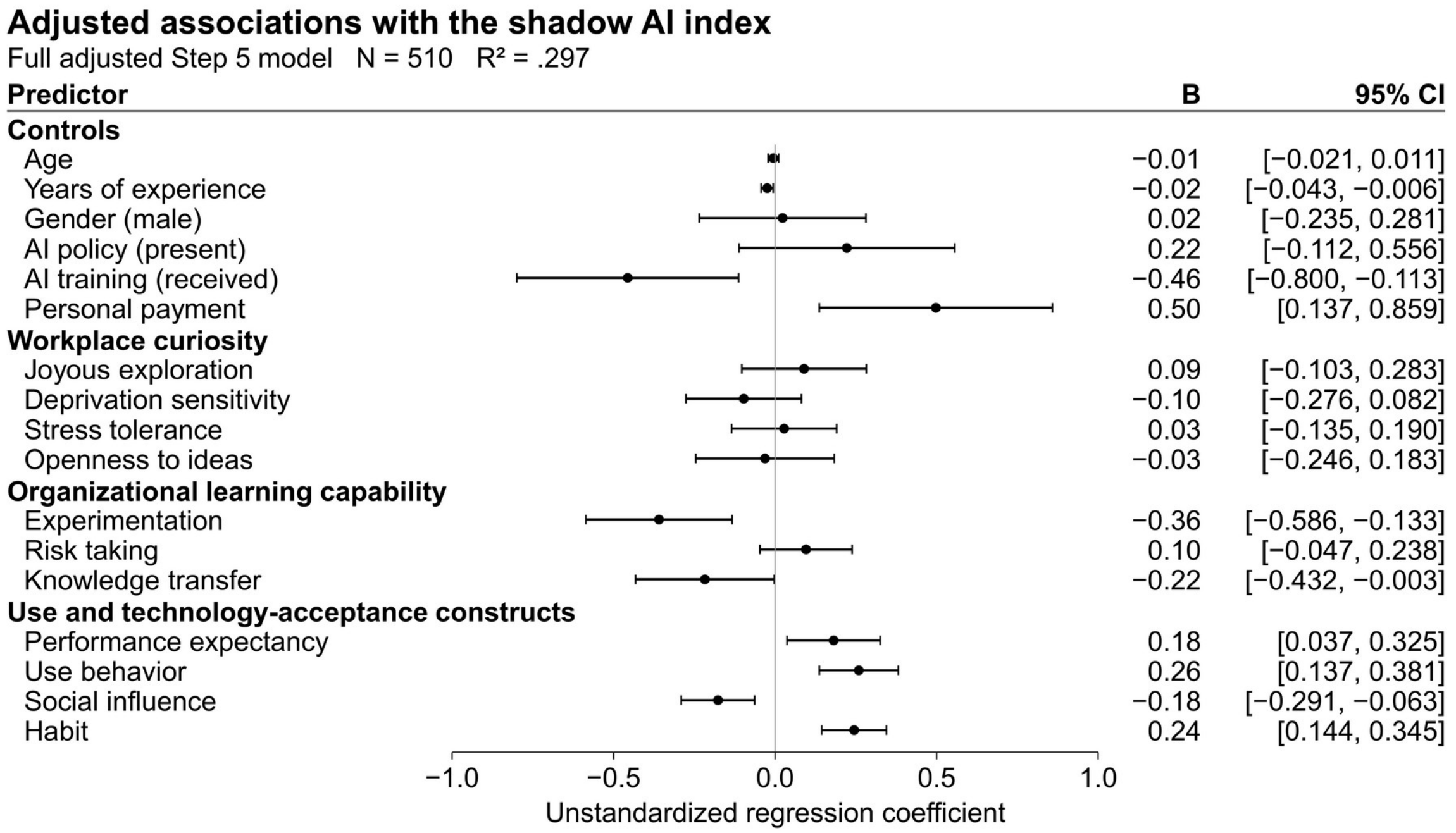

## 5. Discussion

The study compared curiosity with established use patterns and organizational conditions as explanations for shadow AI. Among 11 substantive predictors, joyous exploration fell in the bottom tier, contributing 3.4% of explained variance, while habit contributed 28.3% and completely dominated joyous exploration across all 512 predictor subsets examined. Use behavior captured the frequency of AI engagement; habit captured routinization and felt compulsion. Habit had the largest relative weight and use behavior ranked second. Experimentation and knowledge-transfer conditions were associated with lower shadow AI, and the learning environment collectively accounted for about a quarter of explained variance. In this sample of respondents reporting workplace use during the preceding 12 months, covert use was more strongly associated with routinized behavior and organizational conditions than with exploratory disposition.

### 5.1. Habit and established use

Habit completely dominated joyous exploration across all 512 model subsets, for the composite and for unauthorized use and concealment separately. The distinction matters because curiosity and habit describe different processes: curiosity concerns the motivation to seek novelty, whereas habit concerns behavior that has become routinized through repetition. Habit remained associated with shadow AI after task-specific use frequency was controlled, so its contribution extends beyond how often respondents reported using AI. The habit items concern GenAI use in general, not a directly measured habit of using unauthorized tools or concealing use. They capture self-reported routinization and felt compulsion, and the item-level evidence in Section 4.5 shows associations for both the automaticity item and the compulsion pair. Respondents reporting a stronger felt need for AI also reported more shadow AI. We therefore interpret the finding as an association with self-reported habit strength, which indicates how routinized and necessary AI use feels; automaticity was not observed directly in behavior (Wood & Rünger, 2016).

Limayem et al. (2007) showed that, as behavior becomes more habitual, deliberate intention becomes less predictive of continued use. In the present study, joyous exploration became nonsignificant when performance expectancy, use behavior, and social influence entered together, before habit was added. Habit then explained additional variance, so the attenuation cannot be attributed to habit alone. Polites and Karahanna (2012) similarly showed that incumbent-system habit creates inertia that shapes evaluations of alternatives. The supplementary indirect-association models are consistent with a developmental account in which exploration precedes repeated use and habit, but they do not establish that sequence. Longitudinal evidence is required before habit can be described as a mediator of curiosity's association with shadow AI (Maxwell & Cole, 2007).

Rogers' (2003) adopter categories offer a complementary perspective. The categories classify people by the relative timing of adoption, whereas the present study conditions the focal analysis on employees who had already used generative AI for work during the preceding 12 months and does not observe the timing of initial adoption. Innovators and early adopters may be especially likely to initiate workplace experimentation, while the present findings concern the correlates of more or less covert behavior among respondents with recent workplace use. Personal payment may be consistent with unusually committed or early adoption, but these data cannot assign respondents to adopter categories. Longitudinal research beginning before initial use is needed to distinguish adoption timing from subsequent habit formation.

### 5.2. Concealment as impression management

Treating shadow AI as a human-technology behavior brings concealment into focus as a distinct outcome rather than a byproduct of unauthorized use. The two are correlated but separable ($r = .52$). An impression-management account offers one possible explanation for concealment. Reif et al. (2025) showed that disclosure can carry a social-evaluation penalty, while coworkers who attribute a colleague's AI use to slacking judge that colleague less favorably and offer less help (Zhou et al., 2025). People also judge covert AI use more harshly than open use when directly compared (Purcell et al., 2026), creating competing social risks around disclosure and concealment. The disclosure-factor ratings provide field evidence consistent with the proposed mechanism: anticipated reactions from others uniquely predicted concealment after adjustment for the full model. Because the ratings and outcomes were self-reported concurrently, they do not establish the mechanism. The decomposed models showed a negative coefficient for professional support in concealment, but the paired contrast did not establish a difference from its unauthorized-use coefficient. Together, these patterns direct attention to the perceived social cost of being seen to rely on AI.

### 5.3. Social influence and concealment

Social influence was independently associated with lower shadow AI. Employees who perceived that professionally important others supported AI use reported lower shadow AI ($B = -0.18$, $p = .002$). The concealment coefficient was $B = -0.20$ ($p = .011$), and the unauthorized-use coefficient was $B = -0.12$ ($p = .116$). Their paired difference was nonsignificant ($p = .415$), so the evidence does not establish that professional support relates more strongly to concealment. For the composite outcome, the bivariate association with social influence was near zero ($r = -.02$), whereas the adjusted association was negative. Professional support was positively correlated with both use frequency and habit (Table 2), so the negative coefficient concerns differences in support among employees with similar use patterns and other modeled characteristics. This suppressor pattern should be treated as provisional. Yazdanmehr et al. (2020) showed that social influence operates at organizational and individual levels and can condition security-policy compliance, while Neves and Eisenberger (2014) linked perceived organizational support to employee risk taking through failure-related trust. For shadow AI, professional support may lower the perceived social cost of disclosure, a possibility that warrants direct testing.

### 5.4. The learning environment as additive context

The learning environment was associated with shadow AI independently of the individual predictors. Experimentation and knowledge transfer were associated with lower shadow AI after curiosity, use behavior, and habit were controlled, and the OLC subscales jointly accounted for about a quarter of substantive explained variance, with a larger descriptive share for unauthorized use than concealment. Experimentation may provide legitimate channels for trial and error, while knowledge transfer may reduce uncertainty about which tools and uses are sanctioned. These findings make learning conditions relevant even when employees have similar curiosity, use frequency, and habit scores; the proposed organizational channels themselves were not directly tested. Two findings limit this interpretation. First, OLC did not account for curiosity's attenuation; joyous exploration strengthened slightly when OLC entered and became nonsignificant when performance expectancy, use behavior, and social influence were added, before habit entered. Second, the curiosity-by-OLC interactions were jointly nonsignificant, and nonsignificance does not rule out smaller interactions. The fitted main-effect associations support an additive description, while the presence or absence of smaller interactions remains uncertain.

**5.5. Practical implications**

The findings generate four intervention hypotheses for organizational testing. First, organizations can evaluate sanctioned channels for experimentation, including internal sandboxes, pilot programs, and communities of practice, before unsanctioned routines become established. Habit had the largest relative weight, and experimentation was the most consistent organizational coefficient ($B = -0.36$, $p = .002$), but prospective evaluation is needed to determine whether such channels reduce covert use. Second, personal payment can be treated as a possible provision-gap signal. Employees who pay for their own tools may lack adequate organizational access, prefer different products, or value premium and privacy features. The median expenditure of $20 per month indicates that the observed behavior often involves modest individual spending, but the cross-sectional association does not establish why employees pay.

Third, organizations can evaluate policy and training together prospectively. Respondents reporting policy alone showed the highest descriptive concealment rate (31.2%) and respondents reporting training alone the lowest (13.8%), but the omnibus tests were nonsignificant. The narrower regression result was that organization-provided training was associated with lower shadow AI ($B = -0.46$, $p = .009$). Fourth, organizations can test social-permission interventions. Social influence was associated with lower concealment, and anticipated reactions uniquely predicted concealment in the disclosure-factor analysis. Leadership modeling and team-level discussion may reduce the perceived social cost of open AI use, but experimental or longitudinal research is needed to determine whether these practices change behavior.

### 5.6. Limitations and future directions

**Theoretical delimitations.** The framework does not test the complete UTAUT2 model and does not claim that curiosity is irrelevant to initial adoption. It compares selected UTAUT2 constructs, workplace curiosity, OLC, and governance conditions among employees with recent workplace GenAI use. The data cannot classify Rogers adopter categories or establish a general theory of all shadow AI behavior. The outcome is also provisional. Unauthorized use and concealment are behavioral facets, while risk comfort is attitudinal; separate outcome models and the two-item behavior-only sensitivity analysis reduce, but do not eliminate, this measurement concern. The habit measure blends automaticity with felt compulsion, so the findings concern self-reported habit strength rather than directly observed automaticity.

**Methodological limitations.** All constructs were self-reported, and sensitive conduct may be underreported despite confidentiality protections. The study was not preregistered. Its cross-sectional design permits no causal or temporal inference; indirect associations require longitudinal confirmation (Maxwell & Cole, 2007). Post hoc diagnostics cannot eliminate common method bias. Significant RESET and habit-quadratic checks also indicate that the linear models approximate relationships that may contain curvature. Recruitment used an opt-in Prolific panel, with focal analyses conditioned on workplace GenAI use during the prior 12 months. Without information on nonrespondents or suitable population benchmarks, nonresponse bias cannot be assessed directly. The selection audit compares focal users with screened respondents outside that criterion and does not establish population representativeness. Geographic measurement equivalence was mixed: stress tolerance failed a two-group loading test, the simultaneous loading constraint was significant with a small CFI change, and four-country checks identified a marginal shadow-AI test and inadmissible configural models for joyous exploration, deprivation sensitivity, and risk taking. Small country groups limit those comparisons. Two OLC subscales were excluded for measurement problems. Policy, training, and OLC describe individual perceptions rather than independently verified organizational characteristics; governance cells were unequal. Finally, nonsignificant interaction tests neither establish equivalence nor rule out smaller moderation effects.

**Future research.** Longitudinal designs should begin before initial adoption and follow the transition from exploration to repeated use and habit. Experience-sampling and behavioral measures could separate automaticity from felt dependence and validate unauthorized use and concealment against observable indicators. Experimental studies can test whether sanctioned provision, training, and social-permission interventions reduce covert use. The social-evaluation findings also warrant direct tests of whether anticipated coworker judgments explain concealment. Finally, personal payment should be examined as a bring-your-own-AI phenomenon, with attention to access gaps, product preference, privacy, and task demands.

## 6. Conclusion

Among 11 substantive predictors, joyous exploration contributed about 3% of explained variance and was completely dominated by habit across all 512 predictor subsets examined, for the composite outcome and for unauthorized use and concealment separately. Governance strategies built chiefly around the assumption that curious employees drive unsanctioned use may therefore target a comparatively weak predictor in this sample. The dominant correlates were established use, self-reported habit strength, organizational learning conditions, and professional support for AI use. These findings characterize shadow AI among respondents with workplace use during the preceding 12 months as a problem of routinized behavior and organizational context more than exploratory disposition. The practical question is whether organizations can provide useful, legitimate, and socially acceptable AI pathways before covert routines become established; causal answers require prospective evaluation.

**Table 1**

*Hierarchical Regression Predicting Shadow AI Composite (N = 510)*

| Predictor | *B* | *SE* | *p* |
|---|---|---|---|
| *Step 1: Controls.* $R^2$ = .092 | | | |
| Age | −0.01 | 0.01 | .312 |
| Gender (male) | 0.10 | 0.14 | .472 |
| Years of Experience | −0.02 | 0.01 | .089 |
| AI Policy (present) | 0.21 | 0.19 | .262 |
| AI Training (received) | −0.46* | 0.18 | .012 |
| Personal Payment | 0.92*** | 0.19 | < .001 |
| *Step 2: + Curiosity.* $R^2$ = .118, $\Delta R^2$ = .026, *p* = .006 | | | |
| Joyous Exploration | 0.29** | 0.11 | .008 |
| Deprivation Sensitivity | 0.06 | 0.10 | .540 |
| Stress Tolerance | −0.07 | 0.10 | .436 |
| Openness to Ideas | −0.25* | 0.11 | .031 |
| *Step 3: + Organizational Learning Capability.* $R^2$ = .158, $\Delta R^2$ = .039, *p* < .001 | | | |
| Joyous Exploration | 0.33** | 0.11 | .003 |
| OLC: Experimentation | −0.31* | 0.13 | .014 |
| OLC: Risk Taking | 0.09 | 0.08 | .293 |
| OLC: Knowledge Transfer | −0.24* | 0.12 | .046 |
| *Step 4: + Technology Use Patterns.* $R^2$ = .261, $\Delta R^2$ = .103, *p* < .001 | | | |
| Joyous Exploration | 0.13 | 0.10 | .185 |
| Performance Expectancy | 0.27*** | 0.07 | < .001 |
| Use Behavior | 0.36*** | 0.06 | < .001 |
| Social Influence | −0.16** | 0.06 | .008 |

| Predictor | *B* | *SE* | *p* |
|---|---|---|---|
| *Step 5: + Habit.* $R^2$ = .297, $\Delta R^2$ = .036, *p* < .001 | | | |
| Age | −0.01 | 0.01 | .527 |
| Gender (male) | 0.02 | 0.13 | .860 |
| Years of Experience | −0.02** | 0.01 | .010 |
| AI Policy (present) | 0.22 | 0.17 | .192 |
| AI Training (received) | −0.46** | 0.17 | .009 |
| Personal Payment | 0.50** | 0.18 | .007 |
| Joyous Exploration | 0.09 | 0.10 | .362 |
| Deprivation Sensitivity | −0.10 | 0.09 | .286 |
| Stress Tolerance | 0.03 | 0.08 | .736 |
| Openness to People's Ideas | −0.03 | 0.11 | .774 |
| OLC: Experimentation | −0.36** | 0.12 | .002 |
| OLC: Risk Taking | 0.10 | 0.07 | .188 |
| OLC: Knowledge Transfer | −0.22* | 0.11 | .047 |
| Performance Expectancy | 0.18* | 0.07 | .014 |
| Use Behavior | 0.26*** | 0.06 | < .001 |
| Social Influence | −0.18** | 0.06 | .002 |
| Habit | 0.24*** | 0.05 | < .001 |

*Note.* *N* = 510. HC3 robust standard errors; coefficient *p* values use *t* tests with each model's residual degrees of freedom. Block-change *p* values use conventional nested-model *F* tests. *B* = unstandardized coefficient. All continuous predictors mean-centered. Steps 2 through 4 list newly entered and focal predictors; Step 5 lists the complete final model.

* *p* < .05. ** *p* < .01. *** *p* < .001.

**Table 2**

*Descriptive Statistics and Bivariate Correlations Among Study Variables*

| Variable | *M* | *SD* | 1 | 2 | 3 | 4 | 5 | 6 | 7 | 8 | 9 | 10 | 11 | 12 |
|---|---|---|---|---|---|---|---|---|---|---|---|---|---|---|
| 1. Shadow AI | 2.87 | 1.63 | (.76) | | | | | | | | | | | |
| 2. Joyous Exploration | 3.80 | 0.90 | .12** | (.86) | | | | | | | | | | |
| 3. Deprivation Sensitivity | 3.56 | 0.95 | .08 | .61** | (.87) | | | | | | | | | |
| 4. Stress Tolerance | 3.37 | 0.96 | .04 | .57** | .43** | (.83) | | | | | | | | |
| 5. Openness to People's Ideas | 4.25 | 0.72 | −.06 | .38** | .29** | .40** | (.89) | | | | | | | |
| 6. Habit | 3.41 | 1.61 | .33** | .33** | .30** | .13** | .03 | (.85) | | | | | | |
| 7. Use Behavior | 4.20 | 1.36 | .30** | .47** | .40** | .34** | .18** | .54** | (.84) | | | | | |
| 8. Performance Expectancy | 5.67 | 1.18 | .21** | .45** | .37** | .31** | .26** | .49** | .57** | (.92) | | | | |
| 9. Social Influence | 4.60 | 1.38 | −.02 | .33** | .24** | .27** | .21** | .35** | .46** | .52** | (.94) | | | |
| 10. OLC: Experimentation | 3.45 | 0.90 | −.17** | .36** | .29** | .37** | .35** | .16** | .28** | .34** | .39** | (.85) | | |
| 11. OLC: Risk Taking | 2.91 | 1.06 | −.03 | .31** | .28** | .30** | .18** | .12** | .26** | .22** | .28** | .61** | (.87) | |
| 12. OLC: Knowledge Transfer | 3.94 | 0.77 | −.18** | .31** | .22** | .29** | .40** | .08 | .18** | .28** | .23** | .66** | .40** | (.85) |

*Note.* *N* = 510. Cronbach's alpha reliabilities appear on the diagonal in parentheses. Scales: Shadow AI, Habit, Use Behavior, Performance Expectancy, and Social Influence (1–7); curiosity and OLC subscales (1–5).

** *p* < .01.

**Table 3**

*Relative Weights of Substantive Predictors of Shadow AI With Bootstrap Confidence Intervals*

| Predictor | Epsilon [95% CI] | % of $R^2$ | Rank |
|---|---|---|---|
| Habit | .070 [.037, .109] | 28.3% | 1 |
| Use Behavior | .053 [.029, .085] | 21.7% | 2 |
| OLC: Experimentation | .031 [.013, .056] | 12.5% | 3 |
| Performance Expectancy | .029 [.014, .050] | 11.9% | 4 |
| OLC: Knowledge Transfer | .024 [.010, .049] | 9.8% | 5 |
| Social Influence | .016 [.007, .035] | 6.3% | 6 |
| Joyous Exploration | .008 [.004, .020] | 3.4% | 7 |
| OLC: Risk Taking | .004 [.003, .013] | 1.8% | 8 |
| Openness to People's Ideas | .004 [.002, .016] | 1.5% | 9 |
| Deprivation Sensitivity | .003 [.002, .011] | 1.3% | 10 |
| Stress Tolerance | .003 [.002, .013] | 1.3% | 11 |
| *Grouped* | | | |
| Habit | .070 | 28.3% | |
| OLC (3 subscales) | .059 | 24.1% | |
| Use Behavior | .053 | 21.7% | |
| Performance Expectancy | .029 | 11.9% | |
| Curiosity (4 dimensions) | .018 | 7.5% | |
| Social Influence | .016 | 6.3% | |

*Note*. Relative weights (epsilon) from Johnson (2000); brackets are 95% percentile bootstrap confidence intervals (5,000 resamples; Tonidandel & LeBreton, 2011). Total $R^2$ = .245 (substantive predictors only); percentages use this denominator. General dominance weights also ranked habit first, use behavior second, and joyous exploration seventh (habit 28.0%, use behavior 22.7%, joyous exploration 2.7%). Dominance analysis: habit completely dominates joyous exploration across all 512 possible subsets. Bootstrap intervals for bottom-tier predictors overlap, so exact ranks within that tier are not interpreted.

**Table 4**

*Final-Step Models for Decomposed Shadow AI Outcomes (N = 510)*

| Predictor | Unauthorized use | Concealment | Risk comfort |
|---|---|---|---|
| Age | 0.01 (0.01) | −0.01 (0.01) | −0.01 (0.01) |
| Gender (male) | −0.09 (0.16) | −0.04 (0.17) | 0.20 (0.18) |
| Years of Experience | −0.04** (0.01) | −0.03** (0.01) | −0.00 (0.01) |
| AI Policy (present) | 0.30 (0.19) | 0.42* (0.21) | −0.06 (0.22) |
| AI Training (received) | −0.34 (0.20) | −0.42 (0.22) | −0.61** (0.23) |
| Personal Payment | 0.31 (0.25) | 0.46* (0.24) | 0.72** (0.24) |
| Joyous Exploration | 0.24 (0.13) | 0.04 (0.12) | −0.00 (0.13) |
| Deprivation Sensitivity | −0.22 (0.11) | −0.01 (0.11) | −0.06 (0.12) |
| Stress Tolerance | 0.02 (0.11) | −0.10 (0.11) | 0.16 (0.12) |
| Openness to People's Ideas | −0.14 (0.14) | 0.02 (0.13) | 0.02 (0.14) |
| OLC: Experimentation | −0.51*** (0.15) | −0.17 (0.15) | −0.40** (0.15) |
| OLC: Risk Taking | 0.11 (0.09) | −0.01 (0.11) | 0.18 (0.10) |
| OLC: Knowledge Transfer | −0.13 (0.15) | −0.15 (0.14) | −0.37* (0.15) |
| Performance Expectancy | 0.14 (0.10) | 0.11 (0.10) | 0.29** (0.11) |
| Use Behavior | 0.24** (0.08) | 0.35*** (0.09) | 0.19* (0.08) |
| Social Influence | −0.12 (0.08) | −0.20* (0.08) | −0.21* (0.08) |
| Habit | 0.24*** (0.06) | 0.29*** (0.07) | 0.21** (0.07) |
| $R^2$ | .207 | .236 | .222 |

*Note.* *N* = 510. Entries are unstandardized coefficients (HC3 robust standard errors in parentheses). Stars use two-sided residual-degrees-of-freedom *t* tests. Each model includes the full Step 5 specification (all controls and substantive predictors). Substantive-predictor relative weights: habit accounted for 26.1% (unauthorized use) and 36.2% (concealment) of substantive explained variance and completely dominated joyous exploration across all 512 subsets for both outcomes.

* $p < .05$. ** $p < .01$. *** $p < .001$.

**Table 5**

*Shadow AI by Policy and Training Configuration*

| Configuration | *n* | % unauthorized | % concealment | *M* risk | *M* security |
|---|---|---|---|---|---|
| Neither | 268 | 21.3 | 20.1 | 3.60 | 5.52 |
| Policy only | 48 | 33.3 | 31.2 | 3.58 | 6.02 |
| Training only | 65 | 20.0 | 13.8 | 2.91 | 5.77 |
| Both | 129 | 20.9 | 24.0 | 2.74 | 5.97 |

*Note*. These groups summarize individual respondents' reports. Omnibus ANOVAs: unauthorized use, $F(3, 506) = 1.36$, $p = .253$; concealment, $F(3, 506) = 1.34$, $p = .262$; risk comfort, $F(3, 506) = 6.26$, $p < .001$; security consciousness, $F(3, 506) = 4.03$, $p = .007$. Endorsement percentages use scores of 5 or higher. Omnibus significance does not establish every pairwise contrast.

## Data availability

A deidentified sharing copy of the quality-screened data, with Prolific IDs, free-text responses, and technical metadata removed, is available from the corresponding author on reasonable request together with the codebook, executed R reproduction package, generated verification outputs, and editable figure source. The package includes the run_all.R entry point, session information, source hashes, and a statistic-to-output map. The restricted raw export and pseudonymized master file are not publicly shared.

## Declaration of competing interest

The authors declare that they have no known competing financial interests or personal relationships that could have appeared to influence the work reported in this paper.

## Funding

This research did not receive any specific grant from funding agencies in the public, commercial, or not-for-profit sectors.

## Declaration of generative AI and AI-assisted technologies in the manuscript preparation process

During the preparation of this work, the authors used OpenAI ChatGPT with an author-developed academic editing workflow to improve organization, clarity, and language. The authors reviewed and edited the output as needed and take full responsibility for the content of the article.

**Supplementary Materials**

Beyond Curiosity: Habit and Organizational Learning Capability as Predictors of Shadow AI

These materials provide the construct map, participant-flow and selection audit, reproducibility map and data-sharing protections, final-model confidence intervals, CFA fit statistics and item loadings, and measurement-equivalence checks referenced in the manuscript.

**Table S1**

*Construct Conceptualization, Operationalization, and Analysis Map*

| Construct or condition | Conceptual role | Operationalization | Source or status | Study question | Primary analysis |
|---|---|---|---|---|---|
| Shadow-AI index | Provisional focal outcome | Mean of unauthorized use, concealment, and risk comfort; 1-7 | Study-developed; α = .76 | RQ1-RQ2 | CFA, hierarchical OLS, relative weights, dominance |
| Unauthorized use | Behavioral facet of shadow AI | Single item on unapproved workplace AI use; 1-7 | Study-developed | RQ1-RQ2 | Separate full-model OLS, relative weights, dominance |
| Concealment | Behavioral visibility-management facet | Single item on strategies to avoid detection; 1-7 | Study-developed | RQ1-RQ2 | Separate full-model OLS, relative weights, dominance |
| Risk comfort | Attitudinal ancillary facet | Single item on comfort with unsanctioned AI risk; 1-7 | Study-developed | RQ1-RQ2 | Separate full-model OLS; behavior-only sensitivity excludes it |
| Joyous exploration | Exploratory disposition | Three M-WCS items; 1-5 | Kashdan, Goodman, et al. (2020) | H4a, H7 | Bivariate association, hierarchical OLS, relative weights, dominance, exploratory indirect associations |
| Other curiosity dimensions | Alternative curiosity facets | Deprivation sensitivity, stress tolerance, openness to people's ideas; three items each; 1-5 | M-WCS | H4b, H6 | Hierarchical OLS and relative weights |
| Use behavior | Reported task-specific AI-use frequency | Five task-frequency items; 1-7 | Study-developed task-frequency battery | RQ2, H7 | Hierarchical OLS and relative weights |
| Habit | Routinization and felt dependence | Three UTAUT2 habit items; 1-7 | Venkatesh et al. (2012) | H1, H7 | Hierarchical OLS, relative weights, dominance, item-level robustness |
| Performance expectancy | Instrumental value of AI use | Three UTAUT2 items; 1-7 | Venkatesh et al. (2012) | RQ2 | Hierarchical OLS and relative weights |

| Construct or condition | Conceptual role | Operationalization | Source or status | Study question | Primary analysis |
|---|---|---|---|---|---|
| Social influence | Perceived professional support for AI use | Three UTAUT2 items; 1-7 | Venkatesh et al. (2012) | H2 | Hierarchical OLS and decomposed-outcome models |
| Organizational learning capability | Perceived learning-supportive context | Experimentation, risk taking, and knowledge transfer subscales; 1-5 | Chiva et al. (2007); Tohidi et al. (2012) | H5-H6 | Hierarchical OLS, grouped relative weights, moderation probe |
| Policy, training, and personal payment | Governance and provision conditions | Policy status, organizational training, and personally funded AI access | Study survey controls and condition indicators | H3, RQ3 | Controls, descriptive configurations, payer comparisons |

**Table S2**

*Participant Flow and Sample-Selection Audit*

*Panel A. Participant flow*

| Stage | *N* | Treatment |
|---|---|---|
| Raw Qualtrics response records | 570 | Starting pool |
| Consent decline | 1 | Excluded before quality screening |
| Failed one or both embedded attention checks | 12 | Excluded |
| Additional records below one third of the duration median | 6 | Excluded |
| Quality-screened responses | 551 | All passed both attention checks |
| Did not report workplace GenAI use in the preceding 12 months | 41 | Outside the focal workplace-user sample |
| Focal workplace GenAI users | 510 | Primary analytic sample; complete data on study variables |

**Selection audit.** The available data permit comparison of focal workplace users with screened respondents who did not meet the workplace-use criterion. These comparisons describe selection into the focal analytic sample. They do not estimate population nonresponse or coverage bias because the study lacks suitable nonrespondent information and population benchmarks. Residence data were observed for 506 of 510 focal users and 40 of 41 screened respondents outside the focal criterion; the country comparison uses those observed denominators.

**Table S2 (continued)**

*Participant Flow and Sample-Selection Audit*

*Panel B. Sample-selection audit*

| Variable | Focal users ($n$ = 510) | Screened respondents outside focal use criterion ($n$ = 41) | Standardized difference | $p$ | Test |
|---|---|---|---|---|---|
| Age, years | 39.24 (10.24) | 39.85 (10.51) | −0.06 | .720 | Welch $t$ |
| Professional experience, years | 11.11 (8.51) | 12.15 (9.15) | −0.12 | .488 | Welch $t$ |
| Female | 198/510 (38.8%) | 19/41 (46.3%) | −0.15 | .406 | Fisher exact |
| United States or United Kingdom | 369/506 (72.9%) | 28/40 (70.0%) | 0.06 | .713 | Fisher exact |

| Variable | Focal users ($n$ = 510) | Screened respondents outside focal use criterion ($n$ = 41) | Standardized difference | $p$ | Test |
|---|---|---|---|---|---|
| Formal organizational AI policy | 177/510 (34.7%) | 4/41 (9.8%) | 0.63 | < .001 | Fisher exact |
| Organization-provided GenAI training | 194/510 (38.0%) | 2/41 (4.9%) | 0.88 | < .001 | Fisher exact |
| Self-initiated GenAI training | 255/510 (50.0%) | 13/41 (31.7%) | 0.38 | .034 | Fisher exact |

*Note.* Standardized differences equal the focal-minus-comparison mean or proportion divided by the square root of the average of the two group variances: $(M_1 - M_0)/\text{sqrt}[(s_1^2 + s_0^2)/2]$. For binary indicators, each variance is $p(1 - p)$. Negative values indicate a lower focal-user mean or proportion. These eligibility comparisons are not a nonresponse-bias test.

# S3. Reproducibility map and data availability

Table S3 documents the contents and execution of the reproduction package available from the corresponding author on reasonable request. The listed data and analysis files are not included as journal supplementary files. Paths are relative to the REPRODUCIBILITY folder unless otherwise stated. The native R release was executed; the listed CSV files are generated outputs.

Table S3

Reproduction Package Contents and Execution

| File | Role | Population or content | Execution or use |
|---|---|---|---|
| data/shadow_ai_data_deidentified.csv | Permitted sharing input | 551 screened records; AI.use identifies 510 focal users. Includes country-observed mask and reduced grouping variables. | Excludes IDs and free-text responses; use the supplied codebook. |
| run_all.R | Executed R entry point | Runs the preserved baseline, companion analyses, repair audits, and validation. | From REPRODUCIBILITY, run Rscript run_all.R. |
| analysis/shadow_ai_analysis.R; analysis/chb_revision_analyses.R; analysis/cmv_diagnostics.R | Preserved source scripts | Original analysis specification and corrected June companion. | Retained unchanged; executed through documented wrappers. |
| analysis/wrappers/*.R | Executed adapters and audits | Baseline, June companion, reporting repairs, measurement, diagnostic, selection, and output validation. | Uses relative paths and fixed seeds; inspect all reported warnings. |

Table S3 (continued)

Reproduction Package Contents and Execution

| File | Role | Population or content | Execution or use |
|---|---|---|---|
| outputs_r/*.csv; outputs_r/sessionInfo.txt | Generated results and environment | Model estimates, relative weights, dominance, selection, measurement, governance, diagnostics, and environment. | Generated by the R entry point; source for Tables S2, S4, and S5a–S5d. |

| File | Role | Population or content | Execution or use |
|---|---|---|---|
| STATISTIC_OUTPUT_MAP.csv; HASH_MANIFEST.csv | Result map and file integrity | Links reported quantities to outputs; records package-file hashes. | Check the result map and manifest before reuse. |
| README.md | Execution guide | Dependencies, input privacy boundary, source/adaptor distinction, and known measurement limitations. | Read before reproduction. |
| CHB_D_26_04935_Figures_Editable.pptx | Editable figure source | Conceptual model and full Step 5 adjusted coefficient plot. | Available from the corresponding author on reasonable request; not attached as a separate journal supplementary file. |

**Participant privacy and sharing boundary.** Prolific did not disclose participants' identities to the researchers and supplied pseudonymized platform IDs. The external Qualtrics survey also included optional follow-up contact fields for respondents who volunteered for further contact; those fields and technical metadata remain confined to the restricted raw export and were excluded from the analytic data. The sharing dataset removes Prolific IDs, free-text responses, and technical metadata that could increase re-identification risk.

The restricted raw export and pseudonymized master file are not publicly shared.

**Table S4**

*Final Model With HC3 95% Confidence Intervals*

| Predictor | *B* | HC3 *SE* | *t* | 95% CI | *p* |
|---|---|---|---|---|---|
| Age | −0.005 | 0.008 | −0.633 | [−0.021, 0.011] | .527 |
| Years of experience | −0.024 | 0.009 | −2.588 | [−0.043, −0.006] | .010 |
| Gender (male) | 0.023 | 0.131 | 0.176 | [−0.235, 0.281] | .860 |
| AI policy (present) | 0.222 | 0.170 | 1.306 | [−0.112, 0.556] | .192 |
| AI training (received) | −0.456 | 0.175 | −2.612 | [−0.800, −0.113] | .009 |
| Personal payment | 0.498 | 0.184 | 2.711 | [0.137, 0.859] | .007 |
| Joyous exploration | 0.090 | 0.098 | 0.913 | [−0.103, 0.283] | .362 |
| Deprivation sensitivity | −0.097 | 0.091 | −1.068 | [−0.276, 0.082] | .286 |
| Stress tolerance | 0.028 | 0.083 | 0.338 | [−0.135, 0.190] | .736 |
| Openness to people'*s* ideas | −0.031 | 0.109 | −0.287 | [−0.246, 0.183] | .774 |
| OLC: Experimentation | −0.359 | 0.115 | −3.115 | [−0.586, −0.133] | .002 |
| OLC: Risk taking | 0.096 | 0.073 | 1.317 | [−0.047, 0.238] | .188 |
| OLC: Knowledge transfer | −0.217 | 0.109 | −1.991 | [−0.432, −0.003] | .047 |
| Performance expectancy | 0.181 | 0.073 | 2.472 | [0.037, 0.325] | .014 |
| Use behavior | 0.259 | 0.062 | 4.170 | [0.137, 0.381] | < .001 |
| Social influence | −0.177 | 0.058 | −3.053 | [−0.291, −0.063] | .002 |
| Habit | 0.245 | 0.051 | 4.800 | [0.144, 0.345] | < .001 |

*Note.* $N = 510$. Coefficients are unstandardized, continuous predictors are mean-centered, and residual degrees of freedom are 492. Tests and 95% confidence intervals use the $t$ distribution with HC3 heteroscedasticity-consistent standard errors. Values are rounded directly from the native R output.

Control coding. Gender was coded 1 for male and 0 for the remaining reported categories; AI policy was coded 1 for policy present and 0 for no or unsure; organizational AI training was coded 1 for received and 0 otherwise; personal payment was coded 1 when respondents selected that they paid personally for an AI tool and 0 otherwise. Continuous predictors were mean-centered for regression analyses.

# S5. Measurement models and country-group equivalence

Table S5a presents the full-sample CFA results. The primary model contains the 33 reflective predictor items. The extended model adds five use-behavior items and three shadow-AI items. The M-WCS models compare the four-dimensional curiosity structure with a one-factor alternative.

**Table S5a**

*Full-Sample Confirmatory Factor Analysis Fit Statistics*

| Model | Estimator | $\chi^2$ | *df* | *p* | CFI | TLI | RMSEA | SRMR |
|---|---|---|---|---|---|---|---|---|
| Primary model<br>10 factors, 33 predictor items | MLR | 792.37 | 450 | < .001 | .964 | .958 | .041 | .041 |
| Extended model<br>12 factors, 41 items | ML | 1400.20 | 713 | < .001 | .948 | .940 | .043 | .043 |
| M-WCS<br>4 factors, 12 items | MLR | 48.86 | 48 | .438 | 1.000 | 1.000 | .007 | .022 |
| M-WCS<br>1 factor, 12 items | MLR | 1084.00 | 54 | < .001 | .628 | .545 | .219 | .130 |

*Note.* $N = 510$. MLR = robust maximum likelihood; ML = conventional maximum likelihood. For MLR models, $\chi^2$ is scaled and CFI, TLI, and RMSEA are robust indices. ML rows report conventional indices. SRMR is the standardized root mean square residual. CFI = comparative fit index; TLI = Tucker–Lewis index; RMSEA = root mean square error of approximation. The *p* values test exact model fit. $\chi^2$ is rounded to two decimal places and fit indices to three; values displayed as 1.000 reflect rounding.

The primary and extended models use different estimators and item sets, so their $\chi^2$ values do not form a nested-model comparison. The method-factor sensitivity analysis and the multigroup comparisons below used ML.

Per-scale comparisons constrain factor loadings to equality relative to a configural model. The observed-residence grouping includes 369 United States/United Kingdom respondents and 137 other respondents (N = 506). The legacy CSV indicator has N = 510 because four unknown residences were coded with the other group. The four-country sensitivity uses the United States (246), United Kingdom (123), Canada (48), and Mexico (36), totaling N = 453.

**Table S5b**

*Per-Scale Tests of Loading Equality Across Country Groups*

| Scale | Legacy<br>*N* = 510<br>*p* | Observed<br>*N* = 506<br>*p* | Four-country<br>*N* = 453<br>*p* | Admissibility or interpretation |
|---|---|---|---|---|
| Joyous exploration | .199 | .134 | Not interpreted | Four-country configural model inadmissible. |
| Deprivation sensitivity | .873 | .817 | Not interpreted | Four-country configural model inadmissible. |
| Stress tolerance | .068 | .037 | .237 | Observed-residence loading equality rejected. |
| Openness to ideas | .114 | .155 | .240 | Models admissible. |

| Scale | Legacy $N$ = 510 $p$ | Observed $N$ = 506 $p$ | Four-country $N$ = 453 $p$ | Admissibility or interpretation |
|---|---|---|---|---|
| Performance expectancy | .787 | .823 | .062 | Models admissible. |
| Social influence | .557 | .568 | .803 | Models admissible. |
| Habit | .209 | .224 | .408 | Models admissible. |
| Experimentation | .366 | .264 | .411 | Models admissible. |
| Risk taking | .170 | .221 | Not interpreted | Four-country configural model inadmissible. |
| Knowledge transfer | .329 | .280 | .388 | Models admissible. |
| Use behavior | .863 | .816 | .930 | Models admissible. |
| Shadow-AI index | .434 | .310 | .048 | Four-country loading equality marginal ($p$ = .048). |

*Note.* The observed-residence stress-tolerance comparison gave $\Delta\chi^2(2) = 6.61$, $p = .037$, and $\Delta$CFI = −.008. These results do not support an unqualified claim that every scale has equivalent loadings across country groups. Four-country configural models for joyous exploration, deprivation sensitivity, and risk taking failed admissibility checks; their nominal comparison tests cannot establish equivalence.

**Table S5c**

*Joint Tests of Loading Equality Across Country Groups*

| Joint 12-factor comparison | $N$ | $\Delta\chi^2$ ($\Delta df$) | $p$ | $\Delta$CFI | Metric-model CFI |
|---|---|---|---|---|---|
| Legacy CSV grouping | 510 | 47.41 (29) | .017 | −0.0014 | 0.934 |
| Observed residence | 506 | 50.64 (29) | .008 | −0.0017 | 0.934 |

*Note.* Joint likelihood-ratio tests reject exact loading equality for both groupings, although CFI declines are small. Per-scale $\Delta df$ equals 2 for the two-group tests (3 for experimentation; 4 for knowledge transfer and use behavior); four-country $\Delta df$ equals 6, 9, and 12, respectively. Reported $p$ values are exploratory and unadjusted for multiple testing. All models converged; convergence alone did not establish admissibility.

**Table S5d**

*Standardized Item Loadings for the Extended 12-Factor ML Model*

| Construct | Item identifier | λ | Questionnaire pages |
|---|---|---|---|
| Joyous exploration | JE_1 | .795 | 16–19 |
| | JE_2 | .829 | |
| | JE_3 | .842 | |
| Deprivation sensitivity | DS_1 | .766 | 16–19 |
| | DS_2 | .825 | |
| | DS_3 | .908 | |
| Stress tolerance | ST_1 | .700 | 16–19 |
| | ST_2 | .853 | |
| | ST_3 | .803 | |
| Openness to people’s ideas | OP_1 | .782 | 16–19 |
| | OP_2 | .877 | |

| Construct | Item identifier | λ | Questionnaire pages |
|---|---|---|---|
| | OP_3 | .901 | |
| Performance expectancy | PE_1 | .897 | 19–20 |
| | PE_2 | .892 | |
| | PE_3 | .872 | |
| Social influence | SI_1 | .912 | 20–21 |
| | SI_2 | .940 | |
| | SI_3 | .881 | |
| Habit | Habit_1 | .709 | 22 |
| | Habit_2 | .856 | |
| | Habit_3 | .906 | |

**Table S5d (continued)**

*Standardized Item Loadings for the Extended 12-Factor ML Model*

| Construct | Item identifier | λ | Questionnaire pages |
|---|---|---|---|
| OLC: Experimentation | Exp_1 | .844 | 29–30 |
| | Exp_2 | .824 | |
| | Exp_3 | .734 | |
| | Exp_4 | .707 | |
| OLC: Risk taking | RS_1 | .888 | 29–30 |
| | RS_2 | .748 | |
| | RS_3 | .867 | |
| OLC: Knowledge transfer | KT_I_1 | .745 | 32 |
| | KT_I_2 | .806 | |
| | KT_I_3 | .790 | |
| | KT_I_4 | .754 | |
| | KT_I_5 | .573 | |
| Use behavior | UB_T_1 | .741 | 24–26 |
| | UB_T_2 | .627 | |
| | UB_T_3 | .822 | |
| | UB_T_4 | .760 | |
| | UB_T_5 | .681 | |
| Shadow-AI index | Approach_1 | .750 | 15–16 |
| | Approach_2 | .697 | |
| | Approach_3 | .717 | |

*Note.* $N = 510$. λ = fully standardized factor loading, rounded to three decimal places. All estimates are from the extended 12-factor, 41-item model estimated with conventional maximum likelihood (ML); they are not the loadings of the primary 10-factor MLR model. Item identifiers are retained from the analysis output. Questionnaire pages

refer to the original printed page numbers in the separate Supplementary Questionnaire; add 1 for the PDF page number.